\documentclass[prd,footinbib,twocolumn,floatfix]{revtex4}

\usepackage{amsfonts,amsmath,amssymb}

\newcommand{\CC}{{\mathbb C}}
\newcommand{\ZZ}{{\mathbb Z}}
\newcommand{\QQ}{{\mathbb Q}}
\newcommand{\NN}{{\mathbb N}}
\newcommand{\cD}{{\mathsf D}}
\newcommand{\cq}{{\mathsf q}}
\newcommand{\cb}{{\mathsf b}}
\newcommand{\cL}{{\mathsf L}}
\newcommand{\cH}{{\mathcal H}}
\newcommand{\Hom}{{\rm Hom}}
\newcommand{\ra}{\rightarrow}
\newcommand{\sqG}{{\sqrt {|G|}}}
\renewcommand{\ker}{{\rm Ker}}
\renewcommand{\Im}{{\rm Im}}
\newcommand{\Ann}{{\rm Ann}}

\begin{document}

\title{Ground-state degeneracy for abelian anyons in the presence of gapped boundaries}
\author{Anton Kapustin}
\affiliation{California Institute of Technology}
\begin{abstract}

Gapped phases with long-range entanglement may admit gapped boundaries. If the boundary is gapped, the ground-state degeneracy is well-defined and can be computed using methods of Topological Quantum Field Theory. We derive a general formula for the ground-state degeneracy for abelian Fractional Quantum Hall phases, including the cases when connected components of the boundary are subdivided into an arbitrary number of segments, with a different boundary condition on each segment, and in the presence of an arbitrary number of boundary domain walls.

\end{abstract}

\maketitle

\section{Introduction}

The simplest Fractional Quantum Hall phases with $\nu=1/p$ are characterized by the presence of chiral gapless edge modes \cite{WenFQHreview,WenQFT}. Being chiral, these modes cannot be gapped by any boundary perturbation. However, more general FQH states may have non-chiral edge states which can be lifted. In such cases both the bulk and the boundary are gapped, and one may pose the question of computing the ground-state degeneracy of such a system.  A macroscopic quantum degeneracy would be of particular interest if it occurred in a system with a simple topology, such as a disk or an annulus, since such a system would be easier to realize experimentally.

The simplest possibility is to impose the same boundary condition along each connected component of the boundary. One may also place quasi-particles both in the bulk and on the boundary. However, one can show that for abelian systems this does not lead to quantum degeneracy unless the geometry is non-planar (and thus hard to realize in practice). More generally, one can consider the situation when each connected component has several segments, with different boundary conditions on each segment and with boundary domain walls separating the segments. A boundary domain wall may be regarded as a quasi-particle sitting at the junction of two different kinds of gapped boundaries. We will see that using boundary domain walls it is possible to create a quantum ground-state degeneracy even in a very simple geometry, like a disk. Examples of this kind were recently given in \cite{BJQ1,BJQ2}; we will explain a general procedure for computing the ground-state degeneracy in abelian FQH systems.

Macroscopic properties of abelian FQH phases are described by abelian Chern-Simons theory \cite{WenFQHreview,WenQFT} which is a 3d TQFT. Gapped boundary conditions correspond to topological boundary conditions in TQFT; for abelian Chern-Simons theory they have been studied in \cite{KS1,KS2, Levin} while a more general theory based on fusion categories was developed in \cite{KitaevKong,FSV}.  Our approach is based on reducing the problem to a problem in 2d TQFT which can then be analyzed using fairly standard methods.

\section{Abelian Chern-Simons theory}

Consider a general abelian Chern-Simons gauge theory with gauge group $T\simeq U(1)^N$ and an action
$$
S=\frac{1}{4\pi}\int_M K_{ij} A^i d A^j.
$$
Here the matrix $K$ is symmetric, non-degenerate and integral. If the diagonal entries of $K$ are even, this theory is topological, otherwise it depends on the choice of spin structure on the quantum level. For our purposes this distinction will be unimportant, since we will be mostly interested in topologically trivial spaces.

Point-like defects (i.e. bulk quasi-particles)  are labeled by elements of a finite abelian group $\cD=\Lambda^*/\Im K$, where $\Lambda$ is the lattice $\Hom(U(1),T)$ (the co-character lattice of $T$), $\Lambda^*$ is the dual lattice $\Hom(T,U(1))$ (the character lattice of $T$), and $K$ is regarded as a homomorphism $\Lambda\ra\Lambda^*$. $\cD$ can be regarded as the group of fractional charges of quasi-particles modulo the charges of electrons. 

Another important quantity is a bilinear form $\cb:\cD\times \cD\ra \QQ/\ZZ$ describing the braiding properties of bulk quasi-particles. It is given by
$$
\cb(d,d')=X_i \left(K^{-1}\right)^{ij} X'_j,
$$
where $X,X'\in\Lambda^*$ are pre-images of $d,d'\in \cD$. This bilinear form is non-degenerate in the sense that if $\cb(d,d')$ vanishes for all $d'$ and fixed $d$, then  $d$ must be trivial. Therefore $\cb$ defines an isomorphism between $\cD$ and the Pontryagin-dual group $\cD^*=\Hom(\cD,U(1))$. Explicitly, $d\in\cD$ maps to $g\in\cD^*$ such that $g(d')=\cb(d,d')$ for all $d'\in\cD$. 

A quasi-particle  at a point $p$ with charge $d\in\cD$ creates a flat gauge field whose holonomy along a small circle around $p$ is 
$$
\oint A^i= 2\pi \left(K^{-1}\right)^{ij} X_j,
$$
where $X\in\Lambda^*$ is the pre-image of $d$, as before. The holonomy is an element of $T$, but a rather special one: if we carry a charge $Y$ around $p$ such that $Y$ is in the sublattice $\Im K$, the braiding phase is trivial. One can therefore think of this holonomy as an element $g$ of the group $\cD^*=\Hom(\cD,U(1))$. Note that $g$ is precisely the element of $\cD^*$ which corresponds to $d$ under the isomorphism $\cb$. 

In many regards abelian Chern-Simons theory resembles a 3d gauge theory with a discrete gauge group $\cD^*$. In the latter theory, there are both electric quasi-particles whose charge takes values in $\cD$ and ``vortex'' quasi-particles labeled by holonomy taking values in $\cD^*$. But in Chern-Simons theory the two kinds of particles are identified. We will see in the next section that after compactification to 2d abelian Chern-Simons theory becomes ordinary 2d gauge theory with gauge group $\cD^*$. 

Elementary topological boundary conditions correspond to Lagrangian subgroups of $\cD$, i.e subgroups $\cL$ such that $\cq$ vanishes when restricted to $\cL$ and $|\cL|={\sqrt {|\cD|}}$. The last condition is equivalent to the following requirement: if $\cb(d,l)=0$ for any $l\in \cL$, then $d\in \cL$. The meaning of $\cL$ is this: its elements label those bulk quasi-particles which can be screened once brought to the boundary. One can interpret this as the presence of a condensate of quasi-particles on the boundary whose charges generate $\cL$. Quasi-particles which cannot be screened by the boundary condensate are labeled by elements of the quotient group $\cD/\cL$. Dually, since we are given an isomorphism $\cD\simeq \cD^*$, we can associate to $\cL\subset\cD$ a subgroup $H\subset \cD^*$. By definition, this is a subgroup under which all quasi-particles with charges in $\cL$ are neutral. That is, $H$ can be interpreted as the part of the gauge group which is unbroken by the condensate. The group $\cD/\cL$ labeling boundary quasi-particles is simply the group of charges for $H$. 

Elementary boundary domain walls between boundary conditions corresponding to Lagrangian subgroups $\cL,\cL'$ are labeled by elements of $\cD/(\cL+\cL')$. The physical interpretation is that at the junction of two boundary conditions one has both condensates whose charges generate the subgroup $\cL+\cL'$, thus conserved charge takes values in $\cD/(\cL+\cL')$. Dually, the unbroken gauge group at the junction is $H_1\cap H_2$, and its group of charges is precisely $\cD/(\cL+\cL')$.

We are interested in the situation when the spatial slice is a compact oriented 2-manifold $\Sigma$ with a non-empty  boundary. Each boundary component $\Gamma_a$  is subdivided into segments $\Gamma_a^1,\ldots,\Gamma_a^{N_a}$. On a segment $\Gamma_a^i$ one picks a Lagrangian subgroup $L_a^i$, while at the junction of two consecutive segments $\Gamma_a^i,\Gamma_a^{i+1}$ one picks an element of $\cD/(\cL_a^i+\cL_a^{i+1})$. We are going to compute the dimension of the space of states $\cH$ of abelian Chern-Simons theory in this situation. It is equal to the ground-state degeneracy of the corresponding FQH system.

\section{Reduction to two dimensions}

The dimension of the vector space $\cH$ is equal to the partition function of 3d Chern-Simons theory on $\Sigma\times S^1$. Hence we can rephrase the problem as follows. Consider the 2d TQFT theory obtained by compactifying the Chern-Simons theory on $S^1$. Each topological boundary condition reduces to a topological brane in this 2d TQFT. Boundary domain walls in 3d reduce to particular boundary-changing operators in the 2d TQFT. The dimension of $\cH$ is equal to the 2d topological correlator on $\Sigma$.

Let us identify the 2d TQFT that one obtains by compactification. One of the three components of each $A_i$ becomes a periodic scalar $\phi_i\sim \phi_i+2\pi $, and the action becomes
$$
S_2=\frac{1}{2\pi} \int_\Sigma K_{ij} \phi^i d A^j.
$$
This action describes a topological gauge theory with a discrete gauge group $G=\cD^*=\Hom(\cD,U(1))=\ker K: T\ra T^*$. To see this, let us dualize the scalars $\phi_i$. This leads to the following action:
$$
S_2'=\int_\Sigma (d\chi_i-K_{ij} A^j)\wedge h^i
$$
where $\chi_i$ is a periodic scalar dual to $\phi^i$ and $h^i$ is a 1-form which serves as a Lagrange multiplier field. The fields transform under gauge transformations as follows:
$$
\chi_i\mapsto \chi_i+K_{ij}\sigma^j,\quad A^i\mapsto A^i + d\sigma^i
$$
where $\sigma^i$ are periodic scalars parameterizing gauge transformations. One can think of $\chi_i$ as a field taking values in the dual torus $T^*$ while $\sigma^i$ can be thought of as taking values in $T$. Locally we can fix the gauge $\chi_i=0$ and are left with constant gauge transformations living in the subgroup $\ker K:T\ra T^*$. Thus the path-integral reduces to a sum over flat connections with holonomy in $\ker K\simeq\cD^*$. Note that the 2d theory is sensitive only to $\cD$, not to the bilinear form $\cb$ which were needed to define the parent 3d theory.

The bulk properties of the 2d gauge theory with a discrete abelian gauge group $G$ are very simple. Classical configurations on a circle are labeled by the holonomy of the discrete gauge field, therefore the quantum space of states  has dimension $|G|$ and can be identified with the group algebra $\CC[G]$. By the usual state-operator correspondence, this space can also be thought of as the space of local operators and therefore has the structure of a commutative and associative algebra. The multiplication is given by the convolution of functions on $G$:
$$
(f_1\circ f_2)(g)=\sum_{g'\in G} f_1(g') f_2(g-g').
$$
The partition function on a closed oriented 2-manifold of genus $g$ is given by
$$
Z_2(\Sigma)=\frac{1}{|G|}\sum_{\gamma:\pi_1(\Sigma)\ra G} 1=|G|^{2g-1}.
$$
Note that the partition function is normalized so that $Z(T^2)=|G|$, in agreement with the fact that the dimension of the state space on $S^1$ is $|G|$. Note also that $\CC[G]$ has a natural orthonormal basis $\{e_g, g\in G\}$ given by $e_g(g')=\delta(g-g')$, such that the structure constants of the algebra are nonnegative integers:
$$
e_{g_1}\circ e_{g_2}=e_{g_1+g_2}.
$$
The operator $e_g$ can be thought of as a dimension reduction of a quasi-particle in the 3d theory. Indeed, inverting the state-operator correspondence, we see that $e_g$ is a prescription to perform a path-integral over flat gauge fields whose holonomy around the insertion point is $g$. Since a quasi-particle in 3d creates precisely such a gauge field, its dimensional reduction must be identified with $e_g$, at least up to a numerical factor. 

Before we determine the normalization factor, we need to address a subtlety in the relation between 3d and 2d TQFTs. It is revealed when we compare the partition functions of the 3d TQFT on $\Sigma\times S^1$ and the 2d TQFT on $\Sigma$. The former is the dimension of the state space of the 3d theory on $\Sigma$, which is $|G|^g$ \cite{WenQFT}, while the latter is $|G|^{2g-1}$. The discrepancy arises from the freedom to add terms to the actions of both TQFTs which are integrals of local expressions and are topological invariants of the manifold (3d and 2d, respectively). In 3d, there are no such terms, but in 2d we can add to the action a term
$$
\frac{\mu}{2\pi} \int_\Sigma R\ {\rm vol}_{\Sigma}=\mu \,\chi(\Sigma),
$$
where $\chi(\Sigma)$ is the Euler characteristic of $\Sigma$. Such a term modifies the partition function by a factor $e^{\mu\chi(\Sigma)}$. Thus our 2d TQFT is really a family of 2d TQFTs parametrized by  $\mu$. Reduction from 3d picks a particular $\mu$ which we can determine by comparing the partition functions. Since $\chi=2-2g$, we see that 
$e^{\mu}=\sqG.$

We can now fix the above-mentioned normalization factor by computing the 2d partition function on $S^2$ with two operator insertions $e_{g_1}$ and $e_{g_2}$. From the 3d viewpoint, this must be the dimension of the state space of abelian Chern-Simons theory on $S^2$ with two quasi-particles with holonomies $g_1$ and $g_2$. Therefore we expect to get $\delta(g_1-g_2)$ (if the holonomies do not add up to zero, the space is zero dimensional, otherwise it is one-dimensional because the two quasi-particles can fuse together and annihilate each other). The 2d path-integral, including the normalization factor $e^{\mu\chi}$, is
$$
Z(g_1,g_2)=\frac{\left(\sqG\right)^2}{|G|} \delta(g_1-g_2)=\delta(g_1-g_2),
$$ 
since there is a unique flat connection with prescribed singularities if $g_1=g_2$ and no flat connections otherwise. Thus the operator $e_g$ is precisely the compactification of an elementary quasi-particle in 3d, without the need for any additional factors.

While in the case of discrete gauge theory one can perform the sum over flat connections in the continuum, one can also use a lattice model. To describe this model, let us pick a triangulation of $\Sigma$, or better yet, a cell decomposition of $\Sigma$, since this allows more flexibility. On each oriented 1-cell we place a variable $g\in G$, so that reversing orientation changes $g\mapsto -g$. Gauge transformations are $G$-valued functions on 0-cells, acting on 1-cell  variables in an obvious manner. The partition function is the sum over all choices of 1-cell variables satisfying the flatness constraints, divided by the order of the gauge group. There is one constraint for each 2-cell $P$, and it says that the signed sum of variables corresponding to all 1-cells in the closure of $P$ vanishes, The constraint ensure the absence of ``lattice vortices.'' Finally, an insertion of a local operator $e_g$ corresponds to modifying the flatness prescription for a particular 2-cell so that 1-cell variables sum to $g$ instead of zero. We will find the lattice formulation useful when we study the partition function in the presence of boundaries and boundary domain walls.

\section{Disk correlators in the 2d TQFT}

Topological boundary conditions in the topological 2d gauge theory with gauge group $G$ are labeled by representations of $G$. A boundary condition corresponding to a representation  $V$ is equivalent to placing a charged particle in the representation $V$ on the boundary. In other words, for every boundary component labeled by $V$ one inserts a factor into the path-integral which is the trace of the holonomy of $G$ along this component in the representation $V$. Elementary boundary conditions correspond to irreducible representations of $G$. It is also easy to describe boundary-changing operators: they are maps between representations which commute with the action of $G$. However, not all 2d boundary conditions arise by dimensional reduction from 3d boundary conditions, and moreover an elementary 3d boundary condition after compactification on a circle becomes 
a sum of elementary 2d boundary conditions. For this reason we focus on a special class of 2d boundary conditions labeled by subgroups of $G$. Namely, for every subgroup $H\subset G$ we may consider a boundary where $G$ is broken down to $H$, and no additional weights in the path-integral. This can be achieved by constraining both $A$ and $\chi$ on the boundary. From a representation-theoretic perspective, such a boundary condition corresponds to a representation of $G$ in the space of functions on $G/H$. Clearly, it is a reducible representation (except in the case $H=G$, in which case it is the trivial representation). In the abelian case that we are discussing, the subgroup $H$ acts trivially on such functions, while the rest of the group acts nontrivially. Thus a particle in such a representation breaks the gauge group down to $H$. 

We focus on these 2d boundary conditions because they are more naturally related to 3d boundary conditions. Namely, via the isomorphism $\cb:\cD\ra \cD^*$ every Lagrangian subgroup $\cL\subset \cD$ gives rise to a subgroup $H\subset G$ such that $|H|={\sqrt {|G|}}$. Note that the converse is not true: a 2d boundary condition corresponding to a subgroup $H$ can be lifted to a 3d boundary condition only if $H$ comes from a Lagrangian subgroup of $\cD$. Nevertheless, since in the 2d computations the form $\cb$ does not play a role, we may leave $H$ arbitrary.

Consider now an oriented 2-manifold $\Sigma$ with a non-empty boundary. Each boundary  component is a circle and may be subdivided into segments so that there is a different boundary condition on each segment. At each junction of two boundary segments there is a boundary-changing operator. In the case of interest for us these boundary-changing operators arise from the boundary domain walls in the 3d theory, but this fact is not important for the time being. Using the factorization properties of 2d TQFT, one can replace each boundary component with a local operator. This local operator can be expanded in our preferred basis $\{e_g\}$, and the expansion coefficients are equal to the disk correlator with an insertion of $e_g$ in the bulk. Thus our problem is reduced to evaluating such disk correlators in the 2d TQFT.

To evaluate disk correlators, let us begin with the case when there are no boundary-changing operators, and accordingly the whole boundary is labeled by a subgroup $H$. As usual, the partition function can be evaluated by summing over all flat connections and dividing by the order of the gauge group. The only difference compared to the no-boundary case is that variables living on 1-cells which lie on the boundary take values in $H$, rather than in $G$, and the gauge transformations at the boundary 0-cells also lie in $H$. We also need to multiply the partition function by the factor $\sqrt {|G|}$, since the Euler characteristic of the disk is $1$. The simplest cell decomposition of the disk depicted on fig. 1(a) yields the following expression for the partition function:
$$
Z_{1a}=\frac{\sqG}{H} \delta(g).
$$
If $H$ corresponds to a Lagrangian subgroup in $\cD$, this is equal to $\delta(g)$, i.e. the 3d space of states is one-dimensional if $g=0$ and zero otherwise. This is the expected result, confirming that the partition function was normalized correctly.

\begin{figure}
\begin{picture}(220,120)
\thicklines
\put(0,20){
\put(50,50){\circle{80}}
\put(30,50){\circle*{3}}
\put(75,45){$H$}
\put(50,50){\circle*{2}}
\put(50,40){$g$}
\put(40,-5){$(a)$}
}
\put(120,20){
\put(50,50){\circle{80}}
\put(30,50){\circle*{3}}
\put(70,50){\circle*{3}}
\put(75,45){$\psi_{12}$}
\put(10,45){$\psi_{21}$}
\put(45,74){$H_1$}
\put(45,17){$H_2$}
\put(50,50){\circle*{2}}
\put(50,40){$g$}
\put(40,-5){$(b)$}
}

\end{picture}

\caption{(a) A cellular decomposition of a disk, with an insertion of a local operator $e_g$ in the interior of the 2-cell. (b) A cellular decomposition of a disk, with an insertion of a local operator $e_g$ in the interior of the 2-cell and two boundary-changing operators.}
\end{figure}
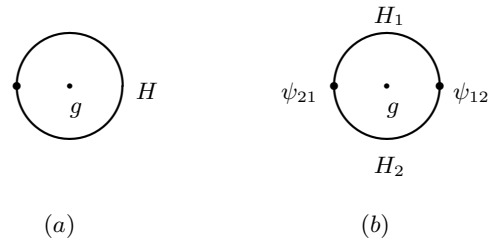

Next we consider the case of a disk with two boundary segments labeled by subgroups $H_1$ and $H_2$, a bulk insertion $e_g$ and two boundary-changing operators $\psi_{12}$ and $\psi_{21}$ (see fig. 1(b)). The space of boundary-changing operators corresponding to a pair of subgroups $H_1,H_2\subset G$ can be identified with the space of functions on the set of double cosets $H_2\backslash G/H_1$, where $H_2$ acts on $G$ ``from the left'', while $H_1$ acts on $G$ ``from the right'':
$$
(h_2,h_1): g\mapsto g+h_2-h_1, \quad h_1\in H_1,\quad h_2\in H_2.
$$
Indeed, by the state-operator correspondence the space of boundary-changing operators can be identified with the space of states of the 2d TQFT on an interval with boundary conditions $H_1$ and $H_2$ on the two ends. Classical configurations are labeled by the holonomy of $G$ modulo gauge transformations, that is, by elements of $H_2\backslash G/H_1$. Hence the quantum space of states is the space of functions on $H_2\backslash G/H_1$. In the abelian case this space is the same as the space of functions on $G$ which are invariant under the action of $H_1+H_2$. Here $H_1+H_2$ is the subgroup of $G$ generated by $H_1$ and $H_2$. In particular, the dimension of the space of boundary-changing operators is
$$
\frac{|G|}{|H_1+H_2|}=\frac{|G| |H_1\cap H_2|}{|H_1| |H_2|}.
$$
A convenient basis in this space is given by functions  
$$
\psi_x(g)=\sum_{g\in G, \pi_{12}(g)=x} e_g,\quad x\in G/(H_1+H_2),
$$ 
where $\pi_{12}:G\ra G/(H_1+H_2)$ is the obvious projection. The function $\psi_x(g)$ is equal to $1$ if $g$ is in the equivalence class of $x$, and zero otherwise. In the path-integral formulation, an insertion of such a boundary operator means that one sums over of flat connections such that the holonomy around a small semi-circle centered at the insertion point is in the equivalence class of $x$, but is otherwise unconstrained. In the lattice model, there is no singularity, of course. Rather, a boundary-changing operator corresponds to a boundary 1-cell separating two boundary segments (see fig. 2). Gauge transformations living on its two end-points take values in $H_1$ and $H_2$. On the 1-cell itself, we have a variable taking values in $G$, but satisfying a constraint that $\pi_{12}(g)=x$.

\begin{figure}
\begin{picture}(50,80)
\put(0,0){
\put(30,10){\line(0,1){80}}
\put(30,60){\circle*{2}}
\put(30,40){\circle*{2}}
\put(35,47){$g\in G$}
\put(17,75){$H_1$}
\put(17,20){$H_2$}
}
\end{picture}

\caption{In the lattice model, a boundary-changing operator $\psi_x$ corresponds to a special 1-cell separating two segments of the boundary. One sums over all values of the variable $g\in G$ living on this 1-cell such that $g$ corresponds to a fixed $x\in G/(H_1+H_2)$.}

\end{figure}
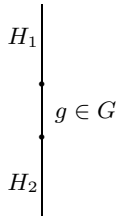

It is fairly clear that up to a normalization factor the operator $\psi_x$ is the dimensional reduction of an elementary boundary domain wall in 3d. One way to argue is as follows. Among all elementary boundary domain walls in 3d there is a distinguished one corresponding to a connection which is nonsingular even at the junction of boundary conditions $\cL_1$ and $\cL_2$. It corresponds to the zero element of $\cD/(\cL_1+\cL_2)$. Clearly, this boundary domain wall reduces to the boundary-changing operator $\psi_0$, perhaps up to a factor. All other elementary boundary domain walls can be obtained by fusing the distinguished one with an elementary bulk quasi-particle labeled by an element $d\in \cD$. Upon compactification, this quasi-particle becomes the bulk operator $e_g$ where $g$ is the image of $d$ under the isomorphism $\cb:\cD\ra \cD^*$. But fusing $e_g$ with $\psi_0$ clearly gives $\psi_x$, where $x=\pi_{12}(g)$. This proves our claim.

To fix the normalization factor, let us compute the partition function corresponding to fig. 1(b). Following the above rules and taking into account only the usual normalization factor $\sqG$, we find
$$
Z_{1b}=\frac{\sqG}{|H_1\cap H_2|}\delta(\pi_{12}(g)+x_{12}+x_{21}).
$$
Note that if $H_1$ and $H_2$ correspond to Lagrangian subgroups in $\cD$, then $|H_1|=|H_2|=\sqG$, and the above partition function is an integer. However, the 3d considerations predict that the dimension of the state space is $1$ if $\pi_{12}(g)+x_{12}+x_{21}=0$ and zero otherwise. Hence in general there is a nontrivial normalization factor relating the compactification of an elementary boundary domain wall and the corresponding basis element $\psi_x$, namely
$$
c_{12}=\sqrt \frac{|H_1\cap H_2|}{\sqG}.
$$
Note that this factor is $1$ in the special case when $H_1=H_2$ and $|H_1|=\sqG$.

We are now ready to compute the partition function on a disk for an arbitrary number of boundary segments and arbitrary boundary-changing operators. Boundary segments are labeled by subgroups $H_1,\ldots, H_N$, while boundary-changing operators are labeled by elements $x_i\in G/(H_i+H_{i+1})$, $i=1,\ldots, N$ with the convention $H_{N+1}=H_1$. Including the normalization factors explained above, we get
\begin{multline*}
Z(x; g)=(\sqG)^{1-N/2} \prod_{i=1}^N \frac{\sqrt {|H_i\cap H_{i+1}|}}{|H_i|}\\
\times \sum_{\begin{array}{c} g_i \in G\\
\pi_{i,i+1}(g_i) =x_i\end{array}} \delta\left(g-\sum_{i=1}^N g_i\right).
\end{multline*}
In the case of interest to us, $|H_i|=\sqG$ for all $i$, so we can simplify this a bit to
\begin{multline*}
Z(x; g)=(\sqG)^{1-3N/2} \prod_{i=1}^N \sqrt {|H_i\cap H_{i+1}|} \\ \times \sum_{\begin{array}{c} g_i \in G\\
\pi_{i,i+1}(g_i) =x_i\end{array}} \delta\left(g-\sum_{i=1}^N g_i\right).
\end{multline*}
It is not at all obvious that this expression is an integer for all conceivable Lagrangian subgroups. We will check this in a few examples below.

 \section{Examples}
 
 Let us compute the ground-state degeneracy in several special cases. First, if $H_1=\ldots=H_N=H$, we get
 $$
 Z(x;g)=\delta\left(\pi_H(g)-\sum_i x_i\right).
 $$
 Here $\pi_H: G\ra G/H$ is the obvious projection.
 Thus if only a single elementary boundary condition is involved, the state space is at most one-dimensional, regardless of the choice of boundary domains walls.
 
 Another simple but interesting case is when two boundary conditions alternate: $H_1$, then $H_2$, then $H_1$ again, etc. Suppose there is a total of $2k$ of boundary segments, $k\in\NN$, so that all boundary domain walls are labeled by elements of the same set $G/(H_1+H_2)$. Then we find:
 $$
 Z(x;g)=\delta\left(\pi_{12}(g)-\sum_i x_i\right) \left(\frac{|H_1|}{|H_1\cap H_2|}\right)^{k-1}.
 $$
In particular, if $H_1\cap H_2=0$ (i.e. $H_1$ and $H_2$ are complementary subgroups of $G$), the degeneracy is $|H_1|^{k-1}$, provided all charges cancel. This agrees with the computation in \cite{BJQ1,BJQ2}, where it was shown that the state space is acted upon by a discrete Weyl algebra of dimension $|H_1|^{2k-2}$.

Finally, let us consider the case of three boundary segments. For simplicity, let us assume that the gauge group $G$ is a product of $H$ and $H^*=\Hom(H,U(1))$, with the obvious pairing $\cb$. That is, $G$ is is a Drinfeld double. One way to construct a Lagrangian subgroup in $H\times H^*$ is to take an arbitrary subgroup $K\subset H$ and let
$$
H_K=K\times \Ann(K),
$$
where $\Ann(K)\subset H^*$ is the annihilator of $K$:
$$
\Ann(K)=\left\{\eta\in H^*\vert \eta\vert_K=0\right\}
$$
Let us pick three subgroups $K_1,K_2,K_3$  in $H$.  Elementary boundary domain walls are labeled by 
$$x_i\in G/(H_{K_i}+H_{K_{i+1}}),\quad i=1,2,3.
$$
Using the identity
$$
\sum_{h_{ij}\in K_i+K_j}\delta(h_{12}+h_{23}+h_{31})=\frac{|K_1||K_2||K_3|}{|K_1\cap K_2\cap K_3|},
$$
we find that the 2d partition function is
$$
Z(x;g)=\delta(\pi_{123}(g-x_1-x_2-x_3)).
$$
where $\pi_{123}$ is the projection to $G/(H_{K_1}+H_{K_2}+H_{K_3})$.
Thus there is no degeneracy in this case.

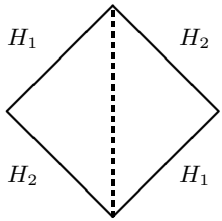
\begin{figure}
\begin{center}
\begin{picture}(100,100)
\put(0,50){
\thicklines
\put(0,0){\line(1,1){40}}
\put(40,40){\line(1,-1){40}}
\put(80,0){\line(-1,-1){40}}
\put(40,-40){\line(-1,1){40}}
\multiput(40,-40)(0,4){20}{\line(0,1){2}}
\put(0,25){$H_1$}\put(0,-27){$H_2$}
\put(65,25){$H_2$}\put(65,-27){$H_1$}
}
\end{picture}
\end{center}
\caption{A nontrivial degeneracy on a triangle can be obtained by starting with a square with an alternating pattern of boundary conditions and folding along the diagonal.}
\end{figure}

One should not think, however, that nontrivial degeneracy is impossible with only three boundary segments. For example, one can take the disk with four boundary segments such there is a nontrivial degeneracy and fold it along a diameter (see fig. 3). This gives a theory with a doubled gauge group on a disk with three boundary segments. All boundary conditions and boundary domain wall are elementary, yet there is a nontrivial degeneracy (the same one as before folding).

\section{Discussion}

We have derived a general and easy-to-use formula for the ground-state degeneracy of abelian FQH systems in the presence of gapped boundaries.
It would be interesting to generalize this computation to nonabelian topological phases in 2+1d systems. A continuum gauge theory description of such phases it not known, in general, but there exists a Hamiltonian lattice model \cite{LevinWen,KitaevKong}. It should be possible to reduce it to 2d and compute the partition function in the resulting 2d TQFT.

\section*{Acknowledgements}

I would like to thank M.~Barkeshli, C.~M.~Jian, and X.~L.~Qi for a discussion. This work was supported in part by the DOE grant DE-FG02-92ER40701.

\end{document}